\documentclass[twocolumn,pra,showpacs,groupedaddress,amsmath,amssymb]{revtex4}

\usepackage{graphicx}
\usepackage{braket}
\usepackage{amsmath}
\usepackage{bm}

\begin{document}

\title{Optimal control of the silicon-based 
donor electron spin quantum computing}
\author{Dong-Bang Tsai} 
\affiliation{Department of Physics and Center for Theoretical Sciences, National Taiwan University, Taipei 10617, Taiwan}
\affiliation{Center for Quantum Science and Engineering, National Taiwan University, Taipei 10617, Taiwan}
\author{Po-Wen Chen} 
\affiliation{Department of Physics and Center for Theoretical Sciences, National Taiwan University, Taipei 10617, Taiwan}
\affiliation{Center for Quantum Science and Engineering, National Taiwan University, Taipei 10617, Taiwan}
\author{Hsi-Sheng Goan} 
	\email{goan@phys.ntu.edu.tw}
\affiliation{Department of Physics and Center for Theoretical Sciences, National Taiwan University, Taipei 10617, Taiwan}
\affiliation{Center for Quantum Science and Engineering, National Taiwan University, Taipei 10617, Taiwan}

\date{\today}
\begin{abstract}
We demonstrate how gradient ascent pulse engineering optimal control methods can be 
implemented on donor electron spin qubits in Si semiconductors with an architecture 
complementary to the original Kane's proposal. We focus on the high-fidelity 
controlled-NOT (CNOT) gate and explicitly find its digitized control sequences by optimizing its fidelity 
over the external controls of the hyperfine $A$ and exchange $J$ interactions. 
This high-fidelity CNOT gate has an error of about $10^{-6}$, below the error
threshold required for fault-tolerant quantum computation, 
and its operation time of $100$ns is about 
$3$ times faster than  $297$ns of the proposed global control scheme.
It also relaxes significantly the stringent distance
constraint of two neighboring donor atoms of $10\sim 20$nm as reported in the 
original Kane's proposal to about $30$nm in which surface $A$ and $J$ gates
may be built with current fabrication technology.
The effects of the control voltage fluctuations, 
the dipole-dipole interaction and the electron spin decoherence  
on the CNOT gate fidelity are also discussed.   
\end{abstract}


\maketitle

One of the important criteria for physical implementation of a 
practical quantum computer is to have a universal set of quantum gates 
with operation times much faster than the relevant decoherence time of 
the quantum computer. In addition, high-fidelity quantum gates
to meet the error threshold of about $10^{-4}$ 
(recently shown to be about $10^{-3}$ in \cite{Aliferis2009})  
are also desired 
for fault-tolerant quantum computation (FTQC). 
There have been several different approaches in optimal control of
quantum gate operation 
problems \cite{Khaneja2005,Yuan2005,Sporl2007}. 
This work focuses on finding
 control parameter sequence 
in near time-optimal way using the gradient
ascent pulse engineering (GRAPE) \cite{Khaneja2005} approach 
for a high-fidelity CNOT gate
in Si:P based 
donor spin quantum computer architectures
\cite{Kane1998,Goan2005,Hill2003,Hill2004} 
where the electron spin 
is defined as qubit \cite{Hill2005}. 
The GRAPE \cite{Khaneja2005} approach
partitions a given time into several equal time steps, and in each time
step of the sequence, the amplitudes of control parameters are 
set to be constant. For a desired operation,
we can define the trace fidelity between the desired operation and
the unitary operation from the sequence. Since we can calculate the
derivative of fidelity with respect to the control amplitudes (gradient
ascent) in each
step, we will be able to obtain, given the required fidelity, 
the near time-optimal control sequence numerically.
Recently, the GRAPE algorithm has been applied to
the coupled Josephson qubit quantum computing \cite{Sporl2007}, and
the numerically optimal control time for a CNOT gate 
is found to be $55$ps \cite{Sporl2007} instead of 
$255$ps in Ref.~\cite{Yamamoto2003}.

The architecture of Si-based donor spin quantum computer 
\cite{Kane1998,Goan2005,Hill2003,Hill2004} 
is composed of $^{31}\mathrm{P}$
atoms doped in a purified $^{28}\mathrm{Si}$ 
host where
each phosphorus has an electron spin and a nuclear spin. 
In a constant magnetic field $B_0$ applied in the $\hat{z}$ direction, the single-qubit Hamiltonian can be written as 
$H=g_{e}\mu_{B}B_{0}\sigma_{z}^{e}/2-g_{n}\mu_{n}B_{0}\sigma_{z}^{n}/2+A\bm{\sigma}^{e}\cdot\bm{\sigma}^{n}$,
 where the effective electron g-factor in Si $g_{e}=2$, the g-factor for a $^{31}P$ nuclear spin $g_{n}=2.26$, 
and the hyperfine interaction 
$A\approx1.21\times10^{-7}\mathrm{eV}$. 
According to numerical calculations \cite{Kettle}, it may be possible to
vary the hyperfine interaction 
with $A$-gate voltage by up to $\approx50\%$ before 
the donor electron is ionized.
Similar to the globally controlled electron spin quantum 
computing scheme \cite{Hill2005}, we apply
a microwave (MW) magnetic field $B_{ac}$ to allow for $x$-axis rotations
and also always
keep the $B_{ac}$ field on 
as it may not be easy to control and turn on/off the $B_{ac}$ field
quickly at the precise times in experiments.
If we initialize the nuclear spins to the spin up state \cite{initialization}, 
we can use the energy states of $\ket{\uparrow_{e}\uparrow_{n}}$ and $\ket{\downarrow_{e}\uparrow_{n}}$
as a qubit \cite{Hill2005}. 
Following Ref.~\cite{Hill2005}, by defining $\omega(A)=\Delta E(A)/\hbar$, 
where
$\Delta E(A)=g_{e}\mu_{B}B_{0}+2A
+[2A^{2}/(g_{e}\mu_{B}B_{0}/2+g_{n}\mu_{n}B_{0}/2)]$,
we obtain the reduced Hamiltonian in the frame rotating with the MW field
\begin{eqnarray}
\tilde{H}={\hbar}\Delta\omega\, \sigma_{z}/2+g_{e}\mu_{B}B_{ac}\sigma_{x}/2,\label{eq:ReduceH1}
\end{eqnarray}
where $\Delta\omega=\omega(A)-\omega_{ac}$, and  $\omega_{ac}$ is the 
angular frequency of the MW field $B_{ac}$.
We tune $\omega_{ac}$ to be the electron spin resonance frequency obtained when no voltage is applied to the corresponding
$A$ gate, i.e., $\omega_{ac}=\omega(A_{0})$. 
Then the qubits will effectively rotate around the $x$-axis 
when $\Delta\omega=0$
(or equivalently $A=A_{0}$), and around an axis which is slightly tilted
when $\Delta\omega\neq0$ (or $A\neq A_{0}$)
described by Eq.~(\ref{eq:ReduceH1}). 


The effective reduced two-qubit Hamiltonian, approximated from
assuming that the nuclear spins are frozen out to be always up, 
in the rotating frame is then
\begin{eqnarray}\raggedright
\tilde{H}=& {\hbar}\Delta\omega_1\sigma^1_{z}/2+{\hbar}\Delta\omega_2\sigma^2_{z}/2+g_{e}\mu_{B}B_{ac}(\sigma^1_{x}+ \sigma^2_{x})/2 \notag\\
&+J\bm{\sigma}^{1e}\cdot\bm{\sigma}^{2e}\;,
\label{eq:ch2:ReduceH}
\end{eqnarray}
where $J$ is the exchange interaction between two adjacent 
donor electron spins. 
We will use the reduced Hamiltonian 
to obtain control sequences
by optimizing the fidelity of 
CNOT gate operations using the GRAPE approach.
Simulations on the full two-qubit Hamiltonian,
\begin{eqnarray}
	H&=& g_{e}\mu_{B}B_{0}(\sigma_{z}^{1e}+\sigma_{z}^{2e})/2-g_{n}\mu_{n}B_{0}(\sigma_{z}^{1n}+\sigma_{z}^{2n})/2 \notag\\
	    & &	        + g_{e}\mu_{B}B_{ac}\cos{\omega_{ac}t}(\sigma_{x}^{1e} +\sigma_{x}^{2e})/2+A_1\bm{\sigma}^{1e}\cdot\bm{\sigma}^{1n} \notag \\
	    &  &			+g_{e}\mu_{B}B_{ac}\sin{\omega_{ac}t}(\sigma_{y}^{1e}+\sigma_{y}^{2e})/2 + A_2\bm{\sigma}^{2e}\cdot\bm{\sigma}^{2n} \notag\\
	    & &  -g_{n}\mu_{n}B_{ac}\cos{\omega_{ac}t}(\sigma_{x}^{1n}+ \sigma_{x}^{2n})/2 + J\bm{\sigma}^{1e}\cdot\bm{\sigma}^{2e}\notag \\
	     & &  -g_{n}\mu_{n}B_{ac}\sin{\omega_{ac}t}(\sigma_{y}^{1n}+\sigma_{y}^{2n})/2  
\;,\label{eq:ch2FullH}
\end{eqnarray}
with the control sequences found will also be performed for error comparison.

Since the $B_{ac}$ field is always on in this scheme,
electrons will undergo a rotation around the $x$-axis when there are
no voltages applied on $A$ gates, i.e. $\Delta\omega=0$, with an angular
frequency of $\Omega_{0}=g_{e}\mu_{B}B_{ac}/\hbar$. While the target
electrons perform a particular unitary operation within time $t$,
every spectator qubit will rotate around the $x$-axis with an angle
of $\theta_{x}=\Omega_{0}t$.
If $\theta_{x}$ does not equal to $2n\pi$  
where $n$ is integral,
another correction step will be required 
for the spectator qubits.
Thus it will be convenient to choose the operation time,
$t  =  2n\pi/\Omega_{0} =  {2n\hbar\pi}/({g_{e}\mu_{B}B_{ac}})$,
such that there is no need for correction for spectator qubits. 
The $B_{ac}$ field is usually 
very small compared with the 
$B_{0}$ field. 
For a given time $t$, we choose $n=1$ 
in the reduced and full Hamiltonian simulations. 
In this case, when the control duration is $100$ns and $n=1$, 
the strength of $B_{ac}$ is $3.56\times10^{-4}\mathrm{T}$.

\begin{figure}
\includegraphics[width=7.0cm]{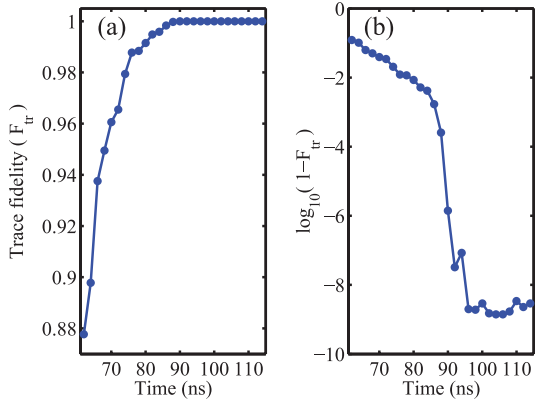} 
\caption{\label{fig:CNOTControlTvsF} (Color online)
Fidelity versus time for the CNOT gate. (a) gives the trace fidelity against
time, while (b) shows deviation $\log_{10}({1-F_{tr}})$ from fidelity.}
\end{figure}
\begin{figure}
\includegraphics[width=7.0cm]{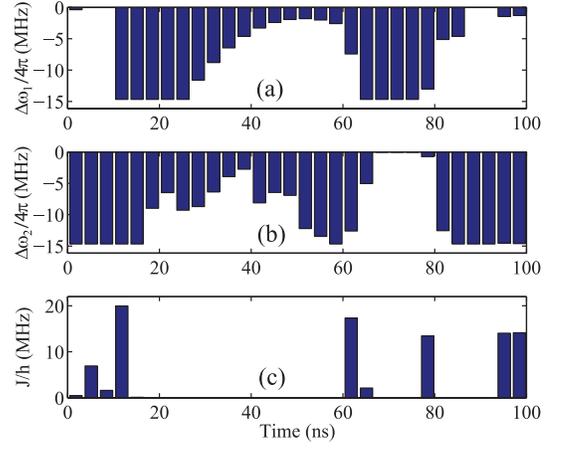} 
\caption{\label{fig:CNOTControlSequence}(Color online)
Near time-optimal CNOT gate control sequence
with 30 steps in $100$ns obtained using the reduced Hamiltonian.
In (a) and (b), the maximum energy difference
of $\sigma_{z}$ term from detuning the hyperfine interaction 
is $(1/2)\Delta\omega/2\pi=-14.7\mathrm{MHz}$.
In (c), the maximum electron-electron exchange energy is $J/h=\mathrm{19.96MHz}$.}
\end{figure}

 
We first try different piecewise constant control steps  
and numerically calculate in the GRAPE approach 
the fidelity (error) against the time needed to implement 
a CNOT gate with stopping criteria of error in the optimizer set to $10^{-9}$ 
in order to economize the simulation time. 
Here, the error is defined as $1-F_{tr}$, where $F_{tr}$ is the trace fidelity defined as $F_{tr}=|\mathrm{Tr}\{U_{D}^{\dagger}U_{F}\}|^{2}$ with $U_{D}$ being the desired unitary operator in a given time $t$, and $U_{F}$ being the optimal unitary operator constructed by our control sequence.
For each trying value of time $t$, we divide the sequence into $30$ piecewise steps, starting
with each of the 
initial control amplitudes ($A_1$, $A_2$ and $J$ gates; or equivalently 
$\Delta\omega_1$, $\Delta\omega_2$ and $J$) by assigning a random
value to every five steps in time 
and using a cubic spline to fill in the amplitudes
of the intermediate time steps. The values of the control amplitudes
$A_1$ and $A_2$ are varied between $A_0/2$ and $A_0$ \cite{Kettle,Hill2005},
and the value of $J$ is varied between $0$ and $J_0$, 
where $J_0$ is chosen for the donor separation to be around $30$nm.
The fidelity
against time obtained from the optimization of the reduced Hamiltonian Eq.~(\ref{eq:ch2:ReduceH}) is shown in Fig.~\ref{fig:CNOTControlTvsF}.
In Fig.~\ref{fig:CNOTControlTvsF} (b), the error 
is less than $10^{-8}$ for times longer than $100$ns, and it is found that $30$ 
piecewise constant control steps for the CNOT gate operation will be sufficient to 
meet the required fidelity (error) and the performance would not be improved further 
with more steps. With the operation time $t=100$ns and stopping criteria of error set to  
$10^{-16}$, we can find that the 
near time-optimal, high-fidelity CNOT gate control 
sequence has an error of $1.11\times10^{-16}$. 
The digitized sequence of controls is 
 shown in Fig.~\ref{fig:CNOTControlSequence}.
In a typical Kane quantum computer's scheme, the typical value
of $J/h\approx 10.2$GHz,
which requires the separation
between two neighboring donors to be 
about $10 \sim 20$nm \cite{Kane1998}. This sets a 
stringent fabrication condition to fabricate surface 
$A$ and $J$ gates within such a short distance. 
One of the great advantages in our scheme is that
the maximum exchange energy in our simulation 
is only $J/h\approx 20$MHz.
This corresponds to a donor separation around $30$nm 
\cite{Kane1998,Herring1963}.
To fabricate gates of this size is
within reach of current fabrication technology.

 
\begin{figure}
\includegraphics[width=8.5cm]{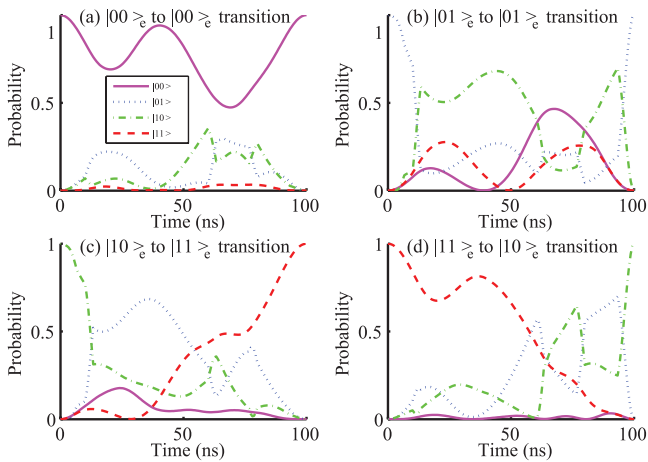}
\caption{\label{fig:CNOTsimulation} (Color online) Time evolution 
of the CNOT gate in the rotating frame, simulated using 
the full Hamiltonian 
with 4 different initial electron-spin input states.
All the nuclear spins are initially spin-up. 
}
\end{figure}
We next apply the control sequence of the CNOT gate, obtained from the optimization of the reduced Hamiltonian Eq.~(\ref{eq:ch2:ReduceH}), to the full spin Hamiltonian Eq.~(\ref{eq:ch2FullH}).
We simulate the CNOT gate numerically with 
initial
four different computational basis electron spin states, 
$\ket{00}_e$, $\ket{01}_e$, $\ket{10}_e$, and $\ket{11}_e$, 
but the same nuclear spin-up state, where 
$\ket{0}_e$ means the electron spin is up. The final reduced electron density matrix is defined 
as the composite density matrix traced over all the nuclear spin states.
The errors of the full Hamiltonian CNOT gate operations with 
the four input electron spin 
basis states evolving
to their correspondingly expected output electron spin states
are shown in Table \ref{tab:CNOTResultE}. 
Here the error is defined as $1-P$, where $P$ is the probability that the qubits are in our desired quantum state after the CNOT operation.
The time evolutions of the states of the CNOT gate 
are shown in Fig.~\ref{fig:CNOTsimulation}.
The error are about $10^{-6}$ which are below the error threshold $10^{-4}$
($10^{-3}$ in \cite{Aliferis2009}) 
required for FTQC. 
Most of the errors result from 
the accuracy of the second-order approximation in $A$ 
of Eq.~(\ref{eq:ch2:ReduceH}) since the hyperfine interaction $A$ 
would cause both electron spins and the nuclear spins to flip 
in the full Hamiltonian (\ref{eq:ch2FullH}).
The CNOT gate operation time of $100$ns is about $3$ times faster than 
the globally controlled electron spin scheme \cite{Hill2005}
of $297$ns [in \cite{Hill2005}, the indicated CNOT time is $148$ns 
that is due to a factor of $2$ missing in the denominator 
of their Hamiltonian Eq.~(6)]. 
The error probabilities that nuclear spins
may flip after the CNOT gate operation for the four input 
electron states are around $10^{-6}$ (see Table \ref{tab:CNOTResultE}).
If we repeat the CNOT process $N$ times by simply inputting the same pure 
electron state $\ket{ij}$ but not reinitializing the nuclear state each time, 
the errors of the CNOT gate operations will 
accumulate. The numerical results indicate that in 
the worst case of the electron spin input state $\ket{10}_e$, after around 
$60$ ($250$) times of operations, 
the error sums up to $1.03\times10^{-4}$ ($0.79\times10^{-3}$). 
Therefore in order to maintain FTQC,
one has to reinitialize the nuclear spin state 
before about $60$ ($250$) times of operations. 
\begin{table}
\caption{\label{tab:CNOTResultE}Summary of the CNOT gate errors.}
\begin{ruledtabular}
\begin{tabular}{cccc}
\footnotesize Input state,  & \footnotesize Expected output         & \footnotesize Error        & \footnotesize Probability that   \\
\footnotesize $\ket{kj}_e\otimes\ket{00}_n$        & \footnotesize state, $\ket{ij}_e\otimes\ket{00}_n$ & ($1-P$)\footnotemark[1]& \footnotesize nuclear spins flip\footnotemark[2]\\
\hline
$\ket{00}_e\otimes\ket{00}_n$ &$\ket{00}_e\otimes\ket{00}_n$ & $1.80\times10^{-8}$ & $1.57\times10^{-7}$\\
$\ket{01}_e\otimes\ket{00}_n$ & $\ket{01}_e\otimes\ket{00}_n$ &$1.80\times10^{-7}$ & $2.00\times10^{-7}$ \\
$\ket{10}_e\otimes\ket{00}_n$ & $\ket{11}_e\otimes\ket{00}_n$ &$1.92\times10^{-6}$ & $1.93\times10^{-6}$\\
$\ket{11}_e\otimes\ket{00}_n$ & $\ket{10}_e\otimes\ket{00}_n$  &$1.20\times10^{-6}$& $1.56\times10^{-6}$\\
\end{tabular}
\footnotetext[1]{The output reduced density matrix of 
the electron spins is obtained by tracing over all the
	 nuclear states.} 
\footnotetext[2]{Here, we trace the total output density matrix over 
the electron spin states to obtain 
the reduced density matrix for the nuclear spin states to 
compute the flipping probability.}
\end{ruledtabular}
\end{table}

Although the exchange interaction dies off exponentially with distance, 
the dipole-dipole interaction that couples every pair of electronic spins in the system
only dies off as $1/d^3$, where $d$ is the distance between two qubits.
The dipole-dipole interaction Hamiltonian can be written as
\begin{eqnarray}
H_D = D\left[\bm{\sigma}^{1e}\cdot\bm{\sigma}^{2e}-3(\bm{\sigma}^{1e}\cdot
      {\hat{n}})(\bm{\sigma}^{2e}\cdot{\hat{n}})\right] \;,
\label{eq:dipole-dipoleH1}
\end{eqnarray}
where $D=\frac{\mu_0\gamma_e^2\hbar^2}{16\pi d^3}$  is the dipolar interaction energy, $\gamma_e=\frac{g_ee}{2m_e}$ is the gyromagnetic ratio of the electrons,
and $\hat{n}$ is the unit vector in the direction joining the two electrons. 
In our scheme, the separation of the two donor 
qubits is around $30$nm, and thus
the corresponding 
$D\approx 1.98\times10^{-12}\mathrm{eV}$, which is still five orders of magnitude smaller than the exchange energy $J$ used in our scheme. 
We simulate the optimal control sequences obtained previously with the full Hamiltonian plus the dipole-dipole interaction Hamiltonian to see its effect. 
Since the first term in Eq.~(\ref{eq:dipole-dipoleH1}) has the same form as the exchange energy, we may combine this term with exchange energy. 
So what we really need to care about is only the second term of Eq.~(\ref{eq:dipole-dipoleH1}),
which becomes $H^{\prime}_D=-3 D {\sigma_y}^{1e}\otimes {\sigma_y}^{2e}$ with the donors aligning along the $\hat{n}=\hat{y}$ axis.
The fidelities of the simulation results are slightly worse than 
the case without dipole-dipole interaction, but they are almost the same and 
the errors are still below the error threshold $10^{-4}$ 
($10^{-3}$ in \cite{Aliferis2009}) 
required for FTQC.
So the dipole-dipole interaction may dominate for larger separations, 
but it is still too small to decrease significantly 
the fidelity of the CNOT gate operation. 

\begin{figure}
\includegraphics[width=7.6cm]{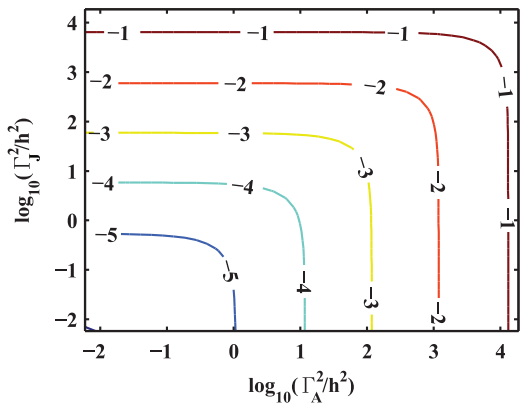}
\caption{\label{fig:GaussNoise} (Color online) 
Contour plot of logarithmic errors
simulated under different spectral densities $\Gamma^2_A$ and $\Gamma^2_J$ of the white noise signals on the control amplitudes of $A$ and $J$ of the full Hamiltonian. The unit of $\Gamma_A^2/h^2$ and $\Gamma_J^2/h^2$ in the plot is $\mathrm{Hz}$ and both of the axes are also in logarithmic scales.}
\end{figure}
Since we apply voltages on the $A$ and $J$ gates to control the strengths
of hyperfine interaction and exchange interaction, there might be
noise induced from the (thermal) fluctuations in the control
circuits, which then cause the uncertainties of the control parameters 
and decrease
the fidelity of a specific operation. To analyze the decrease
of fidelity due to these uncertainties, we model the noise on the 
control parameters $A_1$, $A_2$ and $J$ as independent 
white noise with
Hamiltonian written as 
$H_N =\Gamma_A\xi_1(t)\bm{\sigma}^{1e}\cdot\bm{\sigma}^{1n} 
+ \Gamma_A\xi_2(t)\bm{\sigma}^{2e}\cdot\bm{\sigma}^{2n} 
+ \Gamma_J\xi_3(t)\bm{\sigma}^{1e}\cdot\bm{\sigma}^{2e}$,
where the mean of the continuous time random processes $\langle\xi_i(t)\rangle=0$, the correlation functions 
$\langle\xi_i(t)\xi_j(t^\prime)\rangle=\delta_{ij} \delta(t-t^\prime)$, and 
$\Gamma^2_A$ and $\Gamma^2_J$ are the spectral densities 
of the noise signals, which have the dimension of $\mathrm{(energy)^2/Hz}$.
We simulate the optimal control sequence 
in the presence of the white noise through
the effective master equation approach \cite{Saira2007}.
The contour plot of the logarithmic errors of the 
full-Hamiltonian simulation results due to the white noise 
is shown in Fig.~\ref{fig:GaussNoise}.
To satisfy the error threshold $10^{-4}$ ($10^{-3}$ in \cite{Aliferis2009}) 
of FTQC, the
spectral densities, $\Gamma_J^2/h^2$ and $\Gamma_A^2/h^2$, have
to be smaller than $6.2\mathrm{Hz}$ and $13\mathrm{Hz}$
($63\mathrm{Hz}$ and $125\mathrm{Hz}$), respectively. 
This precision of control should be   
achievable with modern electronic voltage controller devices. 
For example, 
it was stated in \cite{Kane1998} that the spectral density of the gate 
voltage fluctuations for good room temperature electronics is 
of order of $10^{-18}\mathrm{V^2/Hz}$,
comparable to the room temperature Johnson noise of a 
50 $\Omega$ resistor.
At a particular bias voltage, the gates have a 
frequency tuning parameter $\alpha=df/dV$, 
estimated to be $10\sim 100$ MHz/V \cite{Kane1998}.
Therefore, the spectral density of energy fluctuations of 
the control parameters 
for good room temperature devices can be estimated to be
$10^{-4}\sim 10^{-2}\mathrm{Hz}$ 
that is still much smaller than $6\mathrm{Hz}$ required by the
error threshold of $10^{-4}$.  

The decoherence
time $T^e_2$ for P donor electron spin in purified Si 
has been indicated experimentally \cite{Tyryshkin2003}
to be potentially considerably longer than $60$ ms at $4\mathrm{K}$. 
It has been shown \cite{Hill2004} that the two-qubit gate fidelity of Kane's
quantum computer 
is limited primarily by the electron decoherence time, e.g., a
typical error of CNOT is $8.3\times 10^{-5}$ with operation time of 16$\mu$s 
for a simple dephasing model of $T^e_{2}=60 {\rm ms}$. 
In our scheme, the CNOT gate time is much faster and 
we expect the decoherence effect may decrease the fidelity less. 
The error with decoherence  
can be estimated to be $1-F_r e^{-t/T_{2}}$, where $F_r$ and $t$ are 
the trace fidelity and operation time of the gate, respectively.
For this simple estimate, the error is about 
$2.7\times10^{-6}$, below the error threshold of 
$10^{-4}$ ($10^{-3}$ in \cite{Aliferis2009}).

In summary, we have applied the GRAPE approach to find the near time-optimal, high-fidelity CNOT gate control sequence. 
A great advantage of the CNOT gate sequence is that the maximum value of the exchange interaction is $J/h\approx 20$MHz which is  
about $500$ times smaller than the typical value of $10.2$GHz in 
\cite{Kane1998,Hill2003,Hill2004,Hill2005},
and yet the CNOT gate operation time is still about $3$ times faster 
than in \cite{Hill2005}. 
This small exchange interaction 
relaxes significantly the stringent distance constraint of two neighboring 
donor atoms of about $10 \sim 20$nm as reported in the original Kane's 
proposal \cite{Kane1998} to about $30$nm. To fabricate surface gates within such a distance is within reach of current fabrication technology. 
Unlike traditional decomposition method that decomposes general gate operations into several single-qubit and some interaction 
(two-qubit) 
operations in series as the CNOT gate in \cite{Hill2005}, the GRAPE optimal control approach is in a sense more like parallel computing as single-qubit ($A_1$ and $A_2$ both on) and two-qubit ($J$ on) operations can be performed simultaneously on the same qubits in parallel (see Fig.~\ref{fig:CNOTControlSequence}). As a result, the more complex gate operation it is applied, the more time one may save, especially for those multiple-qubit gates that may not be simply decomposed by using the traditional method. So the GRAPE approach may prove useful in implementing quantum gate operations in real quantum computing experiments in the future.
We acknowledge supports from 
NSC under Grant No.~97-2112-M-002-012-MY3, 
from NTU 
under Grants No.~97R0066-65 and 
97R0066-67,
and from the NCTS focus group program. 
We are grateful to NCHC
for computer time and facilities.

\begin{references}
\bibitem{Aliferis2009} P. Aliferis et al., Phys. Rev. A \textbf{79}, 012332 (2009). 
\bibitem{Khaneja2005} N. Khaneja et al., J. Magn. Reson. \textbf{172}, 296 (2005).

\bibitem{Yuan2005} H. Yuan et al., Phys. Rev. A \textbf{72}, 040301(R) (2005); T. Schulte-Herbr\"uggen et al., Phys. Rev. A \textbf{72}, 042331 (2005); 
N. Khaneja et al., Phys. Rev. A \textbf{75}, 012322 (2007);
R. Zeier et al., Phys. Rev. A \textbf{77}, 032332 (2008); 
A. Carlini et al., Phys. Rev. A \textbf{75}, 042308 (2007); S. Montangero et al., Phys. Rev. Lett. \textbf{99} 170501 (2007);
G. Gordon et al., Phys. Rev. Lett. \textbf{101} 010403 (2008).
\bibitem{Sporl2007} A. Sp\"orl et al., Phys. Rev. A \textbf{75} 012302 (2007). 
\bibitem{Kane1998} B. E. Kane, Nature (London) \textbf{393}, 133 (1998); B. E. Kane, Fortsch. Phys.-Prog. Phys. \textbf{48}, 1023 (2000).
\bibitem{Goan2005} H.-S. Goan, Int. J. Quantum Inf. \textbf{3}, 27 Suppl. (2005); L. C. L. Hollenberg et al., Phys. Rev. B \textbf{74} 045311 (2006); C. J. Wellard et al., Phys. Rev. B \textbf{68}, 195209 (2003); B. Koiller et al., Phys. Rev. Lett. \textbf{88}, 027903 (2001); L. M. Kettle et al., Phys. Rev. B \textbf{73}, 115205 (2006).
\bibitem{Hill2003} C. D. Hill et al., Phys. Rev. A \textbf{68}, 012321 (2003).
\bibitem{Hill2004} C. D. Hill et al., Phys. Rev. A \textbf{70}, 022310 (2004).
\bibitem{Hill2005} C. D. Hill et al., Phys. Rev. B \textbf{72}, 045350 (2005).
\bibitem{Yamamoto2003} T. Yamamoto et al.,  Nature (London) \textbf{425}, 941 (2003).
\bibitem{Kettle} L. M. Kettle et al., Phys. Rev. B \textbf{68}, 075317 (2003).
\bibitem{initialization} Our architecture is similar to that 
of \cite{Kane1998}, so we may use the same methods proposed there to initialize the nuclear spins in the spin-up state. Alternatively, we may read out the electron spin state (e.g. as described in \cite{Hill2005}) or wait until the electron spins relax to the spin-down ground state, then apply MW and rf pulse to initialize the electron spins in the spin-up state and then to swap the nuclear and electron spin states. According to our simulations, the error in the initial nuclear spin-up polarization should be kept smaller than $10^{-4}$ (or $10^{-3}$) in order to maintain FTQC.


\bibitem{Herring1963} C. Herring et al., Phys. Rev. \textbf{134}, A362 (1963).

\bibitem{Saira2007} O.-P. Saira et al., Rev. A \textbf{75}, 012308 (2007). 

\bibitem{Tyryshkin2003} S. A. Lyon et al., Rev. B \textbf{68}, 193207 (2003). 

\end{references}


\end{document}